\documentclass [11pt,a4paper] {article}

\usepackage[cp1252]{inputenc}
\usepackage{amssymb}
\usepackage{amsmath}
\usepackage{amsfonts,amssymb}
\usepackage[dvips]{graphicx}
\usepackage{bbm}
\usepackage{enumerate}
\usepackage{amsthm}
\usepackage{cancel}

\DeclareMathAlphabet{\mathpzc}{OT1}{pzc}{m}{it}

\def\SmallColSep{\setlength{\arraycolsep}{1pt}}

\begin{document}

\title{No-cloning implies unalterability of the past}

\author{Arkady Bolotin\footnote{$Email: arkadyv@bgu.ac.il$\vspace{5pt}} \\ \textit{Ben-Gurion University of the Negev, Beersheba (Israel)}}

\maketitle

\begin{abstract}\noindent A common way of stating the non-cloning theorem – one of distinguishing characteristics of quantum theory – is that one cannot make a copy of an arbitrary unknown quantum state. Even though this theorem is an important part of the ongoing discussion of the nature of a quantum state, the role of the theorem in the logical-algebraic approach to quantum theory has not yet been systematically studied. According to the standard point of view (which is in line with the logical tradition), quantum cloning amounts to two classical rules of inference, namely, monotonicity and idempotency of entailment. One can conclude then that the whole of quantum theory should be described through a logic wherein these rules do not hold, which is linear logic. However, in accordance with a supervaluational semantics (that allows one to retain all the theorems of classical logic while admitting ``truth-value gaps''), quantum cloning necessitates the permanent loss of the truth values of experimental quantum propositions which violates the unalterability of the past. The present paper demonstrates this.\\

\end{abstract}

\section{Introduction}  

\noindent In the foundation of quantum mechanics, one of the ``no-go'' (i.e., limitative) theorems of fundamental importance is \textit{the no-cloning theorem} (see papers \cite{Park}, \cite{Wootters}, \cite{Scarani}, and \cite{Ortigoso}, just to name a few). In a nutshell, the theorem establishes the following.\\

\noindent Let $|\Psi\rangle$ denote an original, arbitrary unknown quantum state, and let $|\Psi\Psi\rangle$ symbolize the quantum state whose specification doubles the original state. Imagine the evolution\smallskip

\begin{equation}  
   |\Psi\rangle 
   \longmapsto
   |\Psi\Psi\rangle
   \;\;\;\;   
\end{equation}
\smallskip

\noindent and assume that it is capable of being reversed, that is,\smallskip

\begin{equation}  
   |\Psi\Psi\rangle 
   \longmapsto
   |\Psi\rangle
   \;\;\;\;  .
\end{equation}
\smallskip

\noindent Then, the no-cloning theorem asserts that this evolution cannot be unitary, i.e., one that can be expressed in terms of a unitary operator $U$, namely, $U|\Psi\rangle = |\Psi\Psi\rangle$ and $U^{\dagger}|\Psi\Psi\rangle = U^{\dagger}U|\Psi\rangle = |\Psi\rangle$, where $U^{\dagger}$ is the Hermitian adjoint of $U$. Considering that quantum evolutions are unitary, this means that an arbitrary unknown quantum state cannot be copied.\\

\noindent In logical terms, quantum non-clonability might be interpreted as the failure of two classical rules of inference: \textit{monotonicity of entailment} maintaining (according to \cite{Gentzen}) that the validity of the original conclusion inferred from a list of assumptions $\Gamma$ is not changed by adding an assumption $\mathrm{A}$, in sequential calculus symbols,\smallskip

\begin{equation}  
   \Gamma 
   \vdash
   \Gamma
   ,
   \mathrm{A}
   \;\;\;\;  ,
\end{equation}
\smallskip

\noindent where $\vdash$ is read as ``yields'' or ``entails'', and \textit{idempotency of entailment} stating \cite{Gentzen} that it is possible to delete $\mathrm{A}$ within the context $\Gamma, \mathrm{A}$ in order to infer the same conclusion, in symbols,\smallskip

\begin{equation}  
   \Gamma
   ,
   \mathrm{A}
   \vdash
   \Gamma 
   \;\;\;\;  .
\end{equation}
\smallskip

\noindent As a result, quantum no-cloning – and, thus, the whole of quantum mechanics – can be described through a logic that is stripped from abovementioned rules of inference. Such a logic is Girard's \textit{linear logic} \cite{Girard} (in any of its semantics: phase, coherent or categorical one \cite{Seely, Abramsky}).\\

\noindent Whether or not linear logic is more of ``genuine quantum logic'' than quantum logic of Birkhoff and von Neumann \cite{Birkhoff}, the idea of no-cloning might be construed in completely different logical terms.\\

\noindent Particularly, in \textit{supervaluational semantics} \cite{Varzi}, which allows one to retain all the theorems of classical logic while permitting ``truth-value gaps'', that is to say, admitting that experimental quantum propositions might be neither true nor false nor determinately of any third truth value, quantum non-clonability indicates \textit{the fixity of the past}.\\

\noindent The purpose of the present paper is to demonstrate this.\\

\section{Supervaluational semantics}  

\noindent To adopt supervaluationism as a semantics defining the logic of experimental quantum propositions, one can provide the following reason.\\

\noindent Assume (after \cite{Birkhoff} and \cite{Mackey}) that an experimental quantum proposition, let's denote it by $A$, is uniquely represented by a closed linear subspace, say $\mathcal{A}$, of a Hilbert space $\mathcal{H}$ associated with the quantum system. Subsequently, (in agreement with \cite{Redei}) proposition $A$ will be assigned the value of true if the system is in pure state $|\Phi\rangle$ that belongs to subspace $\mathcal{A}$; in other words, vector $|\Phi\rangle$ is an element of $\mathcal{A}$. In the same way, proposition $A$ will be assigned the value of false if vector $|\Phi\rangle$ is not an element of $\mathcal{A}$.\\

\noindent On the other hand, from the propositional logic perspective, the relation ``is an element of'' is \textit{a predicate} describing the property of ``memberhood'', i.e., being a member. So, if one accepts $\in$ as the name for this predicate, then statements, asserting that a certain quantum state of the system, $|\Psi\rangle$, is (or is not) an element of some closed linear subspace $\mathcal{P}$ of $\mathcal{H}$, can be represented as a Boolean-valued function $\in\!(|\Psi\rangle, \mathcal{P})$, namely,\smallskip

\begin{equation}  
   \in
   :
   |\Psi\rangle
   ,
   \mathcal{P}
   \to
   \{
      \text{true}
      ,
      \text{false}
   \}
   \;\;\;\;  ,
\end{equation}
\smallskip

\noindent where variables $|\Psi\rangle$ and $\mathcal{P}$ stand for arbitrary objects, and symbol $\to$ indicates the domain and codomain of the function.\\

\noindent The value of function $\in\!(|\Psi\rangle, \mathcal{P})$ must determine the truth-value of proposition $P$ represented by subspace $\mathcal{P}$. Accordingly, the assignment of truth values to experimental quantum propositions can be expressed with the seemingly unproblematic formula\smallskip

\begin{equation} \label{EXP} 
   \in
   \!
   \left(
      |\Psi\rangle
      ,
      \mathcal{P}
   \right)
   =
   {[\mkern-3.3mu[
      P
   ]\mkern-3.3mu]}_v
   \;\;\;\;  ,
\end{equation}
\smallskip

\noindent where the double-bracket notation is used to denote \textit{a valuation}, that is, a mapping from the set of elementary (atomic) statements (propositions), symbolized by $\{P\}$, to the set of the truth values, true and false (renamed to 1 and 0, respectively):\smallskip

\begin{equation}  
   v
   \!
   :
   \{ P \}
   \to
   \{
      1
      ,
      0
   \}
   \;\;\;\;  .
\end{equation}
\smallskip

\noindent And yet, the problem with formula (\ref{EXP}) is that, in general cases, the property of being a member may be \textit{neither applicable nor inapplicable} to tuple $(|\Psi\rangle, \mathcal{P})$. This is a characteristic of \textit{vagueness} of predicate $\in$, which is specific to the Hilbert space formalism of quantum mechanics.\\

\noindent Insofar as this vagueness is viewed as a real phenomenon, classical semantics is no longer appropriate as a semantics defining the logic of experimental quantum propositions, and supervaluationism (as the ``standard'' theory of vagueness \cite{Keefe, Williams}) is proposed in its place.\\

\noindent Consequently, in borderline cases that constitute the penumbra of predicate $\in$ – i.e., cases wherein function $\in\!(|\Psi\rangle, \mathcal{P})$ is neither true nor false – proposition $P$ has no truth-value at all. Those cases can be described by\smallskip

\begin{equation}  
   \in
   \!
   \left(
      |\Psi\rangle
      ,
      \mathcal{P}
   \right)
   =
   {[\mkern-3.3mu[
      P
   ]\mkern-3.3mu]}_v
   =
   \frac{0}{0}
   \;\;\;\;  ,
\end{equation}
\smallskip

\noindent where $\frac{0}{0}$ symbolizes an indeterminate value.\\

\noindent To make this explanation more tangible, consider closed linear subspaces of the two-dimensional Hilbert space $\mathbb{C}^2$ characterizing \textit{a qubit}, i.e., a two-state quantum system (as a one-half spin particle, e.g., an electron). Such subspaces are the ranges of operators to measure spin along the $Q$-axis, $\mathrm{ran}(\hat{Q}_{\pm})$, that uniquely represent atomic propositions ``The spin of the qubit along a given axis $Q \in \mathbb{R}^3$ is $\pm \frac{\hbar}{2}\,$'', replaced for brevity with letters $Q_{\pm}$.\\

\noindent For example, in the cases where $Q=Z$ and $Q=X$, one has for subspaces $\mathcal{Z}_{\pm}$ and $\mathcal{X}_{\pm}$ representing propositions $Z_{\pm}$ and $X_{\pm}$, respectively, the following expressions:\smallskip

\begin{equation}  
   \mathcal{Z}_{+}
   =
   \mathrm{ran}
   \!
   \left(
      \!
      \hat{Z}_{+}
      \!
   \right)
   =
   \left\{
      \!\left[
         \begingroup\SmallColSep
         \begin{array}{r}
            c \\
            0 
         \end{array}
         \endgroup
      \right]
   \right\}
   \;\;\;\;  ,
\end{equation}

\begin{equation}  
   \mathcal{Z}_{-}
   =
   \mathrm{ran}
   \!
   \left(
      \!
      \hat{Z}_{-}
      \!
   \right)
   =
   \left\{
      \!\left[
         \begingroup\SmallColSep
         \begin{array}{r}
            0 \\
            c 
         \end{array}
         \endgroup
      \right]
   \right\}
   \;\;\;\;  ,
\end{equation}

\begin{equation}  
   \mathcal{X}_{+}
   =
   \mathrm{ran}
   \!
   \left(
      \!
      \hat{X}_{+}
      \!
   \right)
   =
   \left\{
      \!\left[
         \begingroup\SmallColSep
         \begin{array}{r}
            c \\
            c 
         \end{array}
         \endgroup
      \right]
   \right\}
   \;\;\;\;  ,
\end{equation}

\begin{equation}  
   \mathcal{X}_{-}
   =
   \mathrm{ran}
   \!
   \left(
      \!
      \hat{X}_{-}
      \!
   \right)
   =
   \left\{
      \!\left[
         \begingroup\SmallColSep
         \begin{array}{r}
            c \\
           -c 
         \end{array}
         \endgroup
      \right]
   \right\}
   \;\;\;\;  ,
\end{equation}
\smallskip

\noindent where $c$ is a scalar in $\mathbb{R}$.\\

\noindent Suppose that the qubit is prepared in the state described by the normalized vector such as\smallskip

\begin{equation} \label{INI} 
   |\Phi\rangle
   =
   \left[
      \begingroup\SmallColSep
      \begin{array}{r}
         1 \\
         0 
      \end{array}
      \endgroup
   \right]
   \;\;\;\;  .
\end{equation}
\smallskip

\noindent Since $c|\Phi\rangle$ is an element of $\mathcal{Z}_{+}$, vector $|\Phi\rangle$ is a member of subspace $\mathcal{Z}_{+}$. Accordingly, function $\in\!(|\Phi\rangle, \mathcal{Z}_{+})$ must return the value of 1 which entails the truth of proposition $Z_{+}$:\smallskip

\begin{equation}  
   \in
   \!
   \left(
      \left[
         \begingroup\SmallColSep
         \begin{array}{r}
            1 \\
            0 
         \end{array}
         \endgroup
      \right]
      ,
     \left\{
         \!\left[
            \begingroup\SmallColSep
            \begin{array}{r}
               c \\
               0 
            \end{array}
            \endgroup
         \right]
      \right\}
   \right)
   =
   {[\mkern-3.3mu[
      Z_{+}
   ]\mkern-3.3mu]}_v
   =
   1
   \;\;\;\;  .
\end{equation}
\smallskip

\noindent Along the same lines, because $c|\Phi\rangle$ is not an element of $\mathcal{Z}_{-}$, vector $|\Phi\rangle$ is not a member of $\mathcal{Z}_{-}$. Consequently, function $\in\!(|\Phi\rangle, \mathcal{Z}_{-})$ returns the value of 0 entailing the falsity of proposition $Z_{-}$:\smallskip

\begin{equation}  
   \in
   \!
   \left(
      \left[
         \begingroup\SmallColSep
         \begin{array}{r}
            1 \\
            0 
         \end{array}
         \endgroup
      \right]
      ,
     \left\{
         \!\left[
            \begingroup\SmallColSep
            \begin{array}{r}
               0 \\
               c 
            \end{array}
            \endgroup
         \right]
      \right\}
   \right)
   =
   {[\mkern-3.3mu[
      Z_{-}
   ]\mkern-3.3mu]}_v
   =
   0
   \;\;\;\;  .
\end{equation}
\smallskip

\noindent On the other hand, given that $c|\Phi\rangle$ is \textit{a component} of an element of $\mathcal{X}_{\pm}$, explicitly,\smallskip

\begin{equation}  
   \mathcal{X}_{\pm}
   =
   \left\{
      c
      |\Phi\rangle
      \pm
      c\!
      \left[
         \begingroup\SmallColSep
         \begin{array}{r}
            0 \\
            1 
         \end{array}
         \endgroup
      \right]
   \right\}
   \;\;\;\;  ,
\end{equation}
\smallskip

\noindent vector $|\Phi\rangle$ cannot be regarded as a member of $\mathcal{X}_{\pm}$. But neither can this vector be regarded as not a member of $\mathcal{X}_{\pm}$. Hence, subspaces $\mathcal{X}_{+}$ and $\mathcal{X}_{-}$ create \textit{semantic indeterminacy} for predicate $\in$. In agreement with supervaluationism, it implies that in the state, in which propositions $Z_{+}$ and $Z_{-}$ have the truth-value, neither proposition $X_{+}$ nor proposition $X_{-}$ has one:

\begin{equation}  
   \in
   \!
   \left(
      \left[
         \begingroup\SmallColSep
         \begin{array}{r}
            1 \\
            0 
         \end{array}
         \endgroup
      \right]
      ,
     \left\{
         \!\left[
            \begingroup\SmallColSep
            \begin{array}{r}
                      c \\
               \pm c 
            \end{array}
            \endgroup
         \right]
      \right\}
   \right)
   =
   {[\mkern-3.3mu[
      X_{\pm}
   ]\mkern-3.3mu]}_v
   =
   \frac{0}{0}
   \;\;\;\;  .
\end{equation}
\smallskip

\section{State cloning for qubits}  

\noindent Imagine a system containing two qubits, denoted 1 and 2, which interact with each other such  that pure states $|\Psi^{(1)}\rangle$ of qubit 1 are entangled with pure states $|\Psi^{(2)}\rangle$ of qubit 2. Suppose, one wants to copy (i.e., to transfer the specification of) an unknown state symbolized by $|\Psi^{(1)}_{A}\rangle$ to a preselected state symbolized by $|\Psi^{(2)}_{B}\rangle$.\\

\noindent Clearly, had state $|\Psi^{(1)}_{A}\rangle$ been known, one would have simply selected $|\Psi^{(2)}_{B}\rangle$ to be equal to $|\Psi^{(1)}_{A}\rangle$. As to the direct measurement of $|\Psi^{(1)}_{A}\rangle$, it cannot help with cloning because without prior knowledge of $|\Psi^{(1)}_{A}\rangle$ it irreversibly changes the original state.\\

\noindent Thus, one could hope to accomplish the task only if quantum cloning were to be a causal and reversible evolution of the system in a way that system's state after cloning were to be completely determined by system's state before cloning, and vice versa. In symbols, this evolution can be displayed as follows:\smallskip

\begin{equation} \label{EVOL} 
   |\Psi^{(1)}_{A}\rangle
   \otimes
   |\Psi^{(2)}_{B}\rangle
   \mkern10mu
   \longmapsto
   \mkern10mu
   |\Psi^{(1)}_{A}\rangle
   \otimes
   |\Psi^{(2)}_{A}\rangle
   \;\;\;\;  ,
\end{equation}

\begin{equation} \label{EVOL1} 
   |\Psi^{(1)}_{A}\rangle
   \otimes
   |\Psi^{(2)}_{A}\rangle
   \mkern10mu
   \longmapsto
   \mkern10mu
   |\Psi^{(1)}_{A}\rangle
   \otimes
   |\Psi^{(2)}_{B}\rangle
   \;\;\;\;  .
\end{equation}
\smallskip

\noindent Because $|\Psi^{(1)}_{A}\rangle$ and $|\Psi^{(2)}_{B}\rangle$ can be any arbitrary nontrivial (different from 0) vectors of Hilbert space $\mathbb{C}^2$, let us suppose – for the sake of concreteness – that $|\Psi^{(1)}_{A}\rangle$ is the normalized vector\smallskip

\begin{equation}  
   |\Upsilon\rangle
   =
   \frac{1}{\sqrt{2}}
   \left[
      \begingroup\SmallColSep
      \begin{array}{r}
         1 \\
         1 
      \end{array}
      \endgroup
   \right]
   \;\;\;\;  ,
\end{equation}
\smallskip

\noindent while $|\Psi^{(2)}_{B}\rangle$ is preselected to be vector $|\Phi\rangle$ presented in (\ref{INI}).\\

\noindent In accordance with that which has been discussed in the previous section, at the initial moment of cloning, say $t=0$, the truth-value of proposition $Z_{+}$ is determined by predicate $\in$ on tuple $(|\Phi\rangle, \mathcal{Z}_{+})$, which returns the value that is true.\\

\noindent However, if it were to be possible to copy an unknown state specification of qubit 1 to qubit 2, then, after some fixed time, say $T$, the truth-values of proposition $Z_{+}$ would come to be determined by function $\in\!(|\Upsilon\rangle, \mathcal{Z}_{+})$.\\

\noindent Given that $c|\Upsilon\rangle$ is only one component (out of two) of an element of $\mathcal{Z}_{+}$, i.e.,\smallskip

\begin{equation}  
   \mathcal{Z}_{+}
   =
   \left\{
      c
      \mkern2mu
      |\Upsilon\rangle
      +
      c
      \mkern-3mu
      \left[
         \begingroup\SmallColSep
         \begin{array}{r}
            1 \\
           -1 
         \end{array}
         \endgroup
      \right]
   \right\}
   \;\;\;\;  ,
\end{equation}
\smallskip

\noindent vector $|\Upsilon\rangle$ is not either a member or not a member of subspace $\mathcal{Z}_{+}$. This means that function $\in\!(|\Upsilon\rangle, \mathcal{Z}_{+})$ is neither true nor false, and so proposition $Z_{+}$ in state $|\Upsilon\rangle$ must have no truth-value at all. In symbols,\smallskip

\begin{equation}  
   \in
   \!
   \left(
      \frac{1}{\sqrt{2}}
      \!
      \left[
         \begingroup\SmallColSep
         \begin{array}{r}
            1 \\
            1 
         \end{array}
         \endgroup
      \right]
      ,
     \left\{
         \!\left[
            \begingroup\SmallColSep
            \begin{array}{r}
               c \\
               0 
            \end{array}
            \endgroup
         \right]
      \right\}
   \right)
   =
   {[\mkern-3.3mu[
      Z_{+}
   ]\mkern-3.3mu]}_v
   =
   \frac{0}{0}
   \;\;\;\;  .
\end{equation}
\smallskip

\noindent Hence, the result of quantum cloning can be presented as the loss of the truth value of proposition $Z_{+}$:\smallskip

\begin{equation} \label{LOS1} 
   {[\mkern-3.3mu[
      Z_{+}
   ]\mkern-3.3mu]}_v
   =
   1
   \mkern10mu
   \longmapsto
   \mkern10mu
   {[\mkern-3.3mu[
      Z_{+}
   ]\mkern-3.3mu]}_v
   =
   \frac{0}{0}
   \;\;\;\;  .
\end{equation}
\smallskip

\noindent The similar result would be obtained if, instead of (\ref{INI}), the state $|\Psi_{B}^{(2)}\rangle$ was\smallskip

\begin{equation}  
   |\Psi_{B}^{(2)}\rangle
   =
   \left[
      \begingroup\SmallColSep
      \begin{array}{r}
         0 \\
         1 
      \end{array}
      \endgroup
   \right]
   \;\;\;\;  .
\end{equation}
\smallskip

\noindent In that case, one would find\smallskip

\begin{equation} \label{LOS2} 
   {[\mkern-3.3mu[
      Z_{+}
   ]\mkern-3.3mu]}_v
   =
   0
   \mkern10mu
   \longmapsto
   \mkern10mu
   {[\mkern-3.3mu[
      Z_{+}
   ]\mkern-3.3mu]}_v
   =
   \frac{0}{0}
   \;\;\;\;  .
\end{equation}
\smallskip

\noindent Since at the final moment of cloning, i.e., $t=T$, proposition $Z_{+}$ has no truth-value in both cases, it would be impossible, even in principle, to reconstruct by backward evolution (\ref{EVOL1}) the initial truth-value of this proposition; in symbols,\smallskip

\begin{equation} \label{LOSS} 
   {[\mkern-3.3mu[
      Z_{+}
   ]\mkern-3.3mu]}_v
   =
   \frac{0}{0}
   \mkern10mu
   \longmapsto
   \mkern10mu
   {[\mkern-3.3mu[
      Z_{+}
   ]\mkern-3.3mu]}_v
   =
   \mathbf{X}
   \;\;\;\;  ,
\end{equation}
\smallskip

\noindent where $\mathbf{X}$ stands for a random variable that takes in values 1 and 0 with corresponding probabilities $\Pr(\mathbf{X}=1)$ and $\Pr(\mathbf{X}=0)$ (since vectors $\bigl[\begin{smallmatrix}1\\0\end{smallmatrix}\bigr]$ and $\bigl[\begin{smallmatrix}0\\1\end{smallmatrix}\bigr]$ are symmetrical components of $\mathcal{X}_{+}$, i.e., the subspace whose element is vector $|\Upsilon\rangle$, one’s state of knowledge about the initial truth-value of $Z_{+}$ in state $|\Upsilon\rangle$ must be invariant under this symmetry; consequently, probabilities $\Pr(\mathbf{X}=1)$ and $\Pr(\mathbf{X}=0)$ are reasonably expected to be equal).\\

\noindent It is worth to mention that only if unknown vector $|\Psi_{A}^{(1)}\rangle$ were to be orthogonal to $|\Psi_{B}^{(2)}\rangle$, quantum cloning would not be accompanied by the truth value loss.\\

\noindent One can conclude then that presented above copying of an arbitrary unknown quantum state necessitates the permanent loss of the truth values of experimental quantum propositions.\\

\section{Unalterability of the past}  

\noindent However, this means that such copying would violate the unalterability of the past.\\

\noindent To see this, the first thing to notice is that all propositions are \textit{tensed} \cite{Wolterstorff, Freddoso82, Freddoso83}. This means that all propositions contain the temporal relation between what is asserted in them and the time of this assertion. Consider the following proposition: ``The spin of the qubit along the given axis \textit{is} $+\frac{\hbar}{2}\,$''. Clearly, it is a present-tense proposition, so let’s replace it by symbol $P_{\text{present}}$. In a corresponding manner, symbol $P_{\text{past}}$ replaces the past-tense variant of this proposition, i.e., ``The spin of the qubit along the given axis \textit{was} $+\frac{\hbar}{2}\,$'', and $P_{\text{future}}$ takes the place of the future-tense variant of it: ``The spin of the qubit along the given axis \textit{will be} $+\frac{\hbar}{2}\,$''.\\

\noindent The second thing to notice is that present-tense propositions can change their truth-values, i.e., they can be true at some times and false at others. Thus, suppose that the spin of the qubit along the given axis is $+\frac{\hbar}{2}$ just at one moment of time, e.g., $t=0$. Then, one can say that $P_{\text{present}}$ is true only at $t=0$.\\

\noindent In contrast to that, the truth-values of past-tense propositions cannot change since the past cannot change. That is, if ever a past-tense proposition had the value of true, it would be always true. By way of illustration, because the spin of the qubit along the given axis was $+\frac{\hbar}{2}$ once, $P_{\text{past}}$ is always true and never false.\\

\noindent At this juncture, let us return to quantum cloning. Let moment $t=T$ be the moment of the present. Then, expressions (\ref{LOS1}) and (\ref{LOSS}) can be rewritten as\smallskip

\begin{equation}  
   {[\mkern-3.3mu[
      Z_{+\text{past}}
   ]\mkern-3.3mu]}_v
   =
   1
   \mkern10mu
   \longmapsto
   \mkern10mu
   {[\mkern-3.3mu[
      Z_{+\text{present}}
   ]\mkern-3.3mu]}_v
   =
   \frac{0}{0}
   \;\;\;\;  ,
\end{equation}

\begin{equation}  
   {[\mkern-3.3mu[
      Z_{+\text{present}}
   ]\mkern-3.3mu]}_v
   =
   \frac{0}{0}
   \mkern10mu
   \longmapsto
   \mkern10mu
   {[\mkern-3.3mu[
      Z_{+\text{past}}
   ]\mkern-3.3mu]}_v
   =
   \mathbf{X}
   \;\;\;\;  .
\end{equation}
\smallskip

\noindent As it follows from here, the evolution displayed in (\ref{EVOL}) and (\ref{EVOL1}) can bring about the change of the truth-values of past-tense propositions. To be sure, even though $Z_{+\text{past}}$ had the value of true, after copying and uncopying, there is a 50\% chance that $Z_{+\text{past}}$ would be false.\\

\noindent Therefore, as long as the allowed evolution of quantum systems is one that observes the unalterability of the past, an arbitrary unknown quantum state cannot be copied.\\

\section{Black hole evaporation implies the power over the past}  

\noindent The truth values can be interpreted as values that convey information about experimental quantum propositions \cite{Shramko}. In that case, truth value loss could be regarded as information loss.\\

\noindent It is not difficult to see a likeness between the information loss due to quantum clonability and the information loss due to the black hole radiation (also called Hawking radiation), i.e., a radiation predicted to be released by a black hole because of quantum effects near the event horizon \cite{Hawking74, Hawking76}.\\

\noindent Consider a naturally occurring process that starts off with pure state $|\Psi_{t=0}\rangle$, collapsing into a black hole at some initial time ($t=0$), and ends up with state $|\Psi_{t=T}\rangle$, being emitted by the black hole at time $t=T$, in a way that $|\Psi_{t=T}\rangle$ shares the Hilbert space $\mathcal{H}$ with $|\Psi_{t=0}\rangle$; symbolically,\smallskip

\begin{equation} \label{BHP} 
   |\Psi_{t=0}\rangle
   \mkern10mu
   \longmapsto
   \mkern10mu
   |\Psi_{t=T}\rangle
   \;\;\;\;  .
\end{equation}
\smallskip

\noindent Let $|\Psi_{t=0}\rangle$ be a known quantum state such that predicate function $\in\!(|\Psi_{t=0}\rangle, \mathcal{P})$, where $\mathcal{P}$ is the closed linear subspace of $\mathcal{H}$ representing proposition $P$, returns the value which is 1 (or 0). In opposition, given that (under the simplest models of black hole evaporation) state $|\Psi_{t=T}\rangle$ is a mixed state that can be treated as a random, unknown pure quantum state \cite{Bryan}, predicate $\in$ on tuple $(|\Psi_{t=T}\rangle, \mathcal{P})$ must return neither true nor false. Consequently, the process (\ref{BHP}) implies that the truth values of experimental quantum propositions are lost in black holes:\smallskip

\begin{equation}  
   \in
   \!
   \left(
      |\Psi_{t=0}\rangle
      ,
      \mathcal{P}
   \right)
   \mkern-3.3mu
   =
   \mkern-3.3mu
   {[\mkern-3.3mu[
      \mkern3mu
      P
      \mkern3mu
   ]\mkern-3.3mu]}_v
   =
   1
   \;(\text{or }0)
   \mkern13mu
   \longmapsto
   \mkern13mu
   \in
   \!
   \left(
      |\Psi_{t=T}\rangle
      ,
      \mathcal{P}
   \right)
   \mkern-3.3mu
   =
   \mkern-3.3mu
   {[\mkern-3.3mu[
      \mkern3mu
      P
      \mkern3mu
   ]\mkern-3.3mu]}_v
   =
   \frac{0}{0}
   \;\;\;\;  .
\end{equation}
\smallskip

\noindent Assuming a backward evolution of (\ref{BHP}), one infers from here:\\

\begin{equation}  
   {[\mkern-3.3mu[
      \mkern3mu
      P_{\text{past}}
      \mkern3mu
   ]\mkern-3.3mu]}_v
   =
   1
   \;(\text{or }0)
   \mkern10mu
   \longmapsto
   \mkern10mu
   {[\mkern-3.3mu[
      \mkern3mu
      P_{\text{present}}
      \mkern3mu
   ]\mkern-3.3mu]}_v
   =
   \frac{0}{0}
   \mkern10mu
   \longmapsto
   \mkern10mu
   {[\mkern-3.3mu[
      \mkern3mu
      P_{\text{past}}
      \mkern3mu
   ]\mkern-3.3mu]}_v
   =
   \mathbf{X}
   \;\;\;\;  .
\end{equation}
\smallskip

\noindent That is, black holes have the power over the past, namely, they can change the truth (or falsity) of a statement about the past.\\

\noindent Provided that any naturally occurring process is described by the equations of motion that are continuous, causal, and reversible, and for this reason such a process ought to have the unique past, the last inference presents a serious problem and may suggest a revision of either the models of black hole evaporation or the physical assumptions regarding causality and reversibility.\\

\bibliographystyle{References}

\end{document}